\documentclass[letterpaper,twocolumn,fleqn]{article} 

\usepackage{graphicx} 
\usepackage{amsmath} 
\usepackage{amsfonts}
\usepackage{amssymb}  
\usepackage{hyperref}
\usepackage{subfigure}
\usepackage[utf8]{inputenc}

\title{Fast Large-Scale Model-Based Iterative Tomography via Exploiting Mathematical Structure, Hierarchical Optimization, Smart Initialization, and Distributed GPU Computing}

\author{Dinesh Kumar and Jeffrey J. Donatelli, \\ Center for Advanced Mathematics for Energy Research Applications,  \\ Applied Mathematics and Computational Research Division, \\Lawrence Berkeley National Laboratory, Berkeley, CA}

\date{} 

\begin{document} 

\maketitle 

\thispagestyle{empty} 


\begin{abstract}

Model-Based Iterative Reconstruction (MBIR) is important because direct methods, such as Filtered Back-Projection (FBP) can 
introduce significant noise and artifacts in sparse-angle tomography, especially for time-evolving samples.
Although MBIR produces high-quality reconstructions through prior-informed optimization, 
its computational cost has traditionally limited its broader adoption. In previous work, we addressed this limitation by expressing the Radon transform 
and its adjoint using non-uniform fast Fourier transforms (NUFFTs), reducing computational complexity relative to conventional 
projection-based methods. We further accelerated computation by employing a multi-GPU system for parallel processing.

In this work, we further accelerate our Fourier-domain framework, by introducing four main strategies: 
(1) a reformulation of the MBIR forward and adjoint operators that exploits their multi-level Toeplitz structure for efficient Fourier-domain computation; 
(2) an improved initialization strategy that uses back-projected data filtered with a standard ramp filter as the starting estimate; 
(3) a hierarchical multi-resolution reconstruction approach that first solves the problem on coarse grids and progressively 
transitions to finer grids using Lanczos interpolation; and 
(4) a distributed-memory implementation using MPI that enables near-linear scaling on large high-performance computing (HPC) systems.

Together, these innovations significantly reduce iteration counts, improve parallel efficiency, and make high-quality MBIR reconstruction
 practical for large-scale tomographic imaging. These advances open the door to near-real-time MBIR for applications such as in situ, in operando, and time-evolving experiments.
\end{abstract}

\section{Introduction}
\label{sec:intro}
Synchrotron micro- and nano-computed tomography (micro-CT and nano-CT) \cite{ALS832,APS2BM,NSLS2TXM} have become indispensable 
tools across a wide range of scientific fields, including materials science, 
geoscience, and energy research. These techniques enable non-destructive three-dimensional imaging of the 
internal structure of materials with high spatial resolution. Recent advances in synchrotron instrumentation 
and detector technology have significantly increased acquisition speeds, enabling new experimental modalities 
such as time-resolved tomography.

Time-resolved tomography allows researchers to observe the evolution of microstructural processes within materials. 
Examples include crack propagation under mechanical stress \cite{Larson2018}, 
degradation of polymer electrolytes during battery operation \cite{Kulkarni2020}, and
fluid transport through porous media \cite{boutchko_imaging_2012}. 
Capturing these dynamic processes requires rapid acquisition of projection data, 
often under strict dose or time constraints. As a result, the collected datasets frequently contain limited angular sampling
 and increased measurement noise, making accurate image reconstruction significantly more challenging. Another
application that would benefit from MBIR is laminography which is inherently limited-angle with a missing-wedge because of sample geometry \cite{Marcus:22}. 

Model-based iterative reconstruction (MBIR) has emerged as a powerful framework for addressing these challenges.
By incorporating prior knowledge and accurate physical models into the reconstruction process, MBIR can produce 
higher-quality reconstructions from noisy or incomplete data compared to conventional analytical approaches \cite{Venkat2013, Aditya2014, tomocam2024}. However, these benefits come at the cost of substantial computational expense. Traditional MBIR implementations rely 
on ray-tracing methods for evaluations of forward and adjoint projection operators, which makes reconstruction prohibitively slow for 
large-scale or time-resolved tomography experiments.

To address this limitation, the TomoCAM framework \cite{tomocam2024} was developed as a GPU-accelerated platform for 
tomographic reconstruction. TomoCAM implements the Radon transform and its adjoint using 
non-uniform fast Fourier transforms (NUFFTs) \cite{Greengard2004,Fessler2003}, enabling significantly faster computation 
compared to conventional ray-tracing methods. In addition, the framework leverages multi-GPU architectures to parallelize 
the reconstruction process, allowing large datasets to be processed more efficiently.

In this work, we present several strategies that further improve the performance and scalability of the TomoCAM framework. 
The main contributions of this work are:

\begin{itemize}
\item Reformulating the gradient operator using a multi-level Toeplitz structure enables the use of FFT-based fast convolutions
\item A hierarchical multi-resolution optimization strategy that accelerates convergence by progressively refining the reconstruction grid.
\item A filtered backprojection-based initialization that significantly reduces the number of iterations required for convergence.
\item A hybrid distributed MPI and multi-GPU implementation that enables large-scale tomographic reconstruction on modern HPC systems.
\end{itemize}

The code is available under an open-source license at \url{https://github.com/lbl-camera/tomocam}.

\section{Theory and Mathematical Background}
\label{sec:theory}

Tomographic reconstruction is fundamentally based on the inversion of the
Radon transform. We briefly review the mathematical background here and
refer the reader to standard texts for additional details \cite{CCroke_Chap6, kak2001principles}. 
Let  $f: \mathbb{R}^2 \rightarrow \mathbb{R}$. The Radon transform of $f$ is defined by line-integral
\begin{equation}
    R_\theta \,f (t) = \int_{\ell(t, \theta)} f(x, y) d\ell 
\end{equation}
where $\ell(t, \theta)$ denotes the line parameterized by the offset $t$ and  projection angle $\theta$.
The Radon transform is closely related to the
\emph{Central Slice Theorem}, which states that the one-dimensional Fourier
transform of a projection acquired at angle $\theta$ corresponds to a slice
of the two-dimensional Fourier transform of the object $f(x,y)$ taken at the
same angle $\theta$, i.e.,
\begin{equation}
\mathbb{F} R\,f (\omega) = \hat{f} (\omega \cos \theta, \omega \sin \theta)
\end{equation}
where $\hat{f}$ in two-dimensional Fourier transform of $f$, 
and $(\omega \cos \theta, \omega \sin \theta) = (k_x, k_y)$ the Fourier-domain coordinates.

Traditional iterative tomographic reconstruction methods often rely on repeatedly applying 
forward and back-projection operators, which can be interpreted as large implicit linear systems.
In previous work \cite{tomocam2024}, 
we implemented these operators using
non-uniform fast Fourier transforms (NUFFTs), reducing the computational
complexity from $O(N^2)$ to $O(N \log N + M \log_{10} \epsilon^{-1})$, 
where $N$ denotes the number of voxels in target reconstruction, $M$ is the total number of pixels 
in the projection data, and $\epsilon$ is the desired numerical precision.

The forward projection, i.e., the output of the Radon transform can be calculated as
\begin{equation}
    R f = \mathbb{F}^{-1} \mathcal{F} f,
\end{equation}
where $\mathcal{F}$ denotes the NUFFT Type-2 operation that maps the
Cartesian grid in the object domain to a polar grid in the Fourier domain,
and $\mathbb{F}^{-1}$ denotes the one-dimensional inverse FFT applied along
the radial direction for a given projection angle. Similarly, the adjoint operation, commonly referred to
as back-projection, can be expressed as
\begin{equation}
    R^* g = \mathcal{F}^* \mathbb{F} g,
\end{equation}
where $\mathcal{F}^*$ denotes the NUFFT Type-1 operation that maps the polar
grid in the Fourier domain back to the Cartesian grid in the object domain,
and $\mathbb{F}$ denotes the one-dimensional FFT along the radial direction.

When the projection data doesn't satisfy angular Shannon-Nyquist criterion,
regularization $\Phi$ is required. The tomographic reconstruction problem can then be formulated as the following optimization problem:
\begin{equation}
\label{eq:mbir}
    \hat{f} = \arg \min_f
    \frac{1}{2} \| Rf - g \|_2^2 + \lambda \Phi(|\nabla f|).
\end{equation}

Here, $g$ denotes the measured projection data, $\Phi$ is a regularization
function that enforces prior knowledge, and $\lambda$ is a
regularization parameter that balances data fidelity and regularization.
TomoCAM uses an edge-preserving qGGMRF prior for $\Phi$; we refer the reader to
 \cite{Venkat2013, Aditya2014} for details. The formula for 
the iterative step is obtained by differentiating \eqref{eq:mbir}, i.e.,
\begin{equation}
    \label{eq:gradient}
    f^{k+1} = f^k - \frac{1}{L} \left(R^* (R\,f - g) - \lambda \nabla \Phi (|\nabla f|)\right),
\end{equation}
where $L$ is the Lipchitz constant of the gradient.

It is worth noting that, for 3D reconstruction, the Radon-based data fidelity term is evaluated slice-by-slice, 
whereas the qGGMRF prior enforces spatial coupling across the full 3D volume.
This optimization problem can be solved using various iterative algorithms.
In this work, we employ the Fast Iterative Method with
Restart \cite{GiselssonB14b}, an accelerated variant of Nesterov's method \cite{Nesterov2014},
due to its favorable convergence properties for large-scale problems.

In the following sections, we describe several newly implemented strategies that further
improve the convergence rate and computational efficiency of this framework.

\subsection{Toeplitz Structure Exploitation}
\label{sec:toeplitz}

The primary computational bottleneck in solving Eq.~(\ref{eq:mbir}) arises from
repeated evaluations of $R^*(R f - g)$ in \eqref{eq:gradient}, during gradient
computation.  Each evaluation requires two NUFFTs. Although NUFFTs have
log-linear complexity, the gradient calculation can be further accelerated 
by exploiting the Toeplitz structure of operator composition $R^*\,R$.
Specifically, this composition can be expressed as a convolution between $f$ and a
\emph{point spread function} (PSF) kernel $K$, that depends on the coordinates
sampled in the Fourier space. 

While Toeplitz acceleration has been explored in MRI reconstruction \cite{FESSLER2007191, Ou:EECS-2017-90} and  cryo-EM \cite{wang2013fourier}, 
its application to NUFFT based tomographic formulations has received limited attention.
This observation allows the data fidelity gradient to be computed using FFT-based convolution.

The data fidelity gradient is given by
\begin{equation}
   \label{eqn:toeplitz}
    \nabla_f e = R^* (R f - g) = R^* R f - R^* g. 
\end{equation}

In \eqref{eqn:toeplitz}, we can rewrite 
\begin{equation}
    R^* R f = K \circledast f,
\end{equation}
where $\circledast$ refers to convolution.

The PSF kernel $K$ can be precomputed by applying $\mathcal{F}^*$ to a
function $\mathbf{1}$, where $\mathbf{1}(x,y) = 1$ for all $(x,y)$ 
sampled in Fourier space.
To accurately recover $R^* R \,f $, the PSF kernel $K$ must be sampled on a larger grid with at least $2N-1$  grid points in each dimension. Accordingly, to compute the gradient,
$f$ is zero-padded to size $2N-1$  in each dimension, and the resulting convolution is restricted to an $N \times N$ region. The convolution can then be evaluated efficiently using FFTs:
\begin{equation}
    R^* R f =
    Res_{N \times N}(\mathbb{F}^{-1} \left( \mathbb{F}K \odot \mathbb{F} Pad_{2 N - 1 \times 2 N -1}f \right),
\end{equation}
where $\mathbb{F}$ denotes the two-dimensional fast Fourier transform, $Res$ is restriction, $Pad$ is extension, and
$\odot$ represents element-wise multiplication. The reformulated loss and gradient
equations are given as,
\begin{align}
    e &= \frac{1}{2}f^\intercal \, K\circledast f - f^\intercal \, R^*g + \frac{1}{2}g^\intercal g \label{eq:toeploss} \\
    \nabla_f e &= K\circledast f - R^*g \label{eq:toepgrad}
\end{align}

Both $\mathbb{F}K$ and $R^* g$ are computed once prior to the iterative reconstruction. 
This reformulation significantly reduces the computational cost of the data fidelity gradient
and loss, as it requires only two FFTs and one element-wise multiplication instead of
a forward and a backprojection.

\subsection{Initialization Strategy}

The choice of initialization can significantly influence the convergence
behavior of iterative reconstruction algorithms. Starting the
optimization from a poor initial estimate may require many iterations
before the algorithm approaches a meaningful solution. To mitigate this
issue, we adopt an informed initialization strategy based on filtered
backprojection (FBP), which provides a fast approximate reconstruction
from the measured projection data \cite{CCroke_Chap6, kak2001principles}.

Let $g$ denote the measured projection data and $R$ the forward
projection operator. A quick initial estimate can be obtained by
applying the adjoint operator $R^*$ to the projection data. However,
an unfiltered backprojection produces a blurred object due to the
frequency response of the Radon transform. To compensate for this
effect, the projection data are first filtered with a ramp filter prior
to backprojection \cite{RamLak1971}.

Let $p_\theta(s)$ denote the projection at angle $\theta$ and sinogram
coordinate $s$. The filtered projection is computed as
\begin{equation}
\tilde{p}_\theta(s) = (h \circledast p_\theta)(s),
\end{equation}
where $\circledast$ denotes convolution and $h(s)$ is the ramp filter kernel.
In the Fourier domain, the ramp filter is defined as
\begin{equation}
\hat{h}(\omega) = |\omega|,
\end{equation}
which amplifies high-frequency components to compensate for the
smoothing introduced by the projection process.

The initial guess is then obtained using filtered backprojection
\begin{equation}
f_0 = R^* \tilde{g},
\end{equation}
where $\tilde{g}$ denotes the collection of filtered projections.
This initialization provides a computationally inexpensive
approximation of the true solution.

Although the filtered backprojection reconstruction may contain
artifacts due to limited-angle sampling or measurement noise, it
typically captures the dominant low-frequency structure of the
underlying image. Consequently, it serves as an effective starting
point for the iterative optimization. By initializing the reconstruction with $f_0$, 
the algorithm begins closer to the desired
solution, reducing the number of iterations required for convergence
and lowering the overall computational cost.

\subsection{Hierarchical Multi-Resolution Reconstruction}
\label{sec:hierarchical}
To accelerate convergence, we employ a hierarchical multi-resolution
reconstruction strategy. Instead of solving the inverse problem directly
at the target resolution, the reconstruction is performed progressively
across a sequence of grids with increasing spatial resolution. This
approach leverages the observation that the low-frequency components of
the solution dominate the early stages of reconstruction and can be
estimated reliably on coarser grids at significantly lower computational
cost.

We define a hierarchy of grids
$\Omega_0, \Omega_1, \dots, \Omega_L$, where $\Omega_0$ corresponds to the
coarsest discretization and $\Omega_L$ to the desired reconstruction
resolution. At each level $l$, the optimization problem in
Eq.~(\ref{eq:mbir}) becomes
\begin{equation}
f^{(l)} =
\arg\min_f \;
\frac{1}{2}\|R_{(l)} f - g_{(l)}\|_2^2 +
\lambda \Phi(|\nabla f|),
\end{equation}
where $R_{(l)}$ denotes the forward operator defined on grid
$\Omega_l$, and $g_{(l)}$ represents projection data that have been
downsampled to match that resolution. For downsampling, we select 
coarse-grid points by subsampling the full-resolution projection data.

The optimization is first performed on the coarsest grid $\Omega_0$,
producing an estimate $f^{(0)}$. After convergence at this level, the
solution is interpolated to the next finer grid using Lanczos
interpolation. Lanczos interpolation is based on a windowed sinc kernel that provides
high-quality band-limited interpolation while suppressing ringing
artifacts.

Let $P_l$ denote the interpolation operator mapping grid
$\Omega_l$ to $\Omega_{l+1}$. The initialization for the next level is
given by $f^{(l+1)}_0 = P_l f^{(l)}.$
The Lanczos kernel with window parameter $a$ is defined as
\begin{equation}
L(x) =
\begin{cases}
\frac{\sin(\pi x) \, \sin(\pi a x)}{a \pi^2 x^2}, & |x| < a, \\
0, & \text{otherwise},
\end{cases}
\end{equation}

In practice, the interpolation operator $P_l$ is implemented as a
convolution with a separable Lanczos kernel. For a discrete signal
$f[i]$, the interpolated value at location $x$ is given by

\begin{equation}
\tilde{f}(x) = \sum_i f[i]\, L(x-i).
\end{equation}

For multi-dimensional reconstructions, the interpolation is applied
separably along each spatial dimension. The Lanczos window truncates
the sinc kernel to a finite support $|x|<a$, limiting computational cost
while preserving most of the spectral properties of ideal sinc
interpolation. In our implementation, we use a fixed window size $a = 3$,
which provides a good compromise between interpolation accuracy and
computational efficiency.

The reconstruction is then refined by solving the optimization problem
on $\Omega_{l+1}$ using $f^{(l+1)}_0$ as the initial estimate. This
coarse-to-fine procedure is repeated until the finest grid $\Omega_L$
is reached.

This hierarchical strategy accelerates convergence because large-scale
structures are recovered on coarse grids where the number of
unknowns is small and the optimization landscape is smoother. The
interpolated solution therefore provides a high-quality initialization
for subsequent levels, allowing the optimization on finer grids to
focus primarily on recovering high-frequency details. As a result,
high-quality reconstructions can be obtained with significantly fewer
iterations at the finest resolution.

\subsection{Distributed Multi-GPU Implementation}
\label{sec:distributed_gpu}

To handle large-scale tomographic datasets and accelerate computation,
the reconstruction framework is implemented using a hybrid parallel
architecture that combines distributed-memory parallelism with GPU
acceleration. Inter-node communication is handled using the Message
Passing Interface (MPI), while GPUs within each node are coordinated
using shared-memory parallelism. This design distributes both the
computational workload and memory footprint across multiple GPUs
while maintaining efficient evaluation of the gradient and
regularization terms required by the iterative reconstruction algorithm.

The reconstruction volume is
partitioned along the rotation axis of the tomography geometry.
The domain is first decomposed across MPI processes (compute nodes),
and each node further partitions its assigned sub-volume among the
GPUs available on that node. This hierarchical decomposition enables
efficient scaling across both nodes and GPUs.

The data fidelity gradient depends only on local volume partitions and no point-to-point communication between neighboring processes is
required. In contrast,  the qGGMRF prior used in this work relies on interactions between each voxel and its 26 nearest neighbors in
three dimensions. Thus, boundary information must be exchanged
between adjacent sub-volumes during each iteration. To support this,
a halo region is maintained for each partition that stores boundary
voxels from neighboring processes. After every iteration, halo
regions are updated through MPI communication to ensure that the
regularization gradient is computed consistently across partition
boundaries.

Within each compute node, GPU execution is orchestrated using a
lightweight scheduling system designed to overlap data transfers
and GPU computation. The scheduler consists of three cooperating
threads connected through thread-safe work queues. The first thread
transfers input data from host memory to GPU memory and inserts the
corresponding work items into a bounded work queue. The second thread
retrieves tasks from this queue and executes the associated GPU
kernels. Completed tasks
are placed into an outgoing queue, from which a third thread
asynchronously transfers the results from GPU memory back to host
memory.


The resulting hybrid MPI--GPU framework scales efficiently on modern
multi-GPU HPC systems, enabling reconstruction of large volumetric
datasets that would otherwise exceed the memory capacity of a single
device. By combining distributed computation, shared-memory
parallelism, and GPU acceleration, the proposed implementation
makes high-quality model-based iterative reconstruction (MBIR)
practical for large-scale tomographic imaging applications.

\section{Numerical Benchmarks}

\begin{figure}[h!]
 \subfigure[]{
    \centering
    \includegraphics[width=0.21\textwidth]{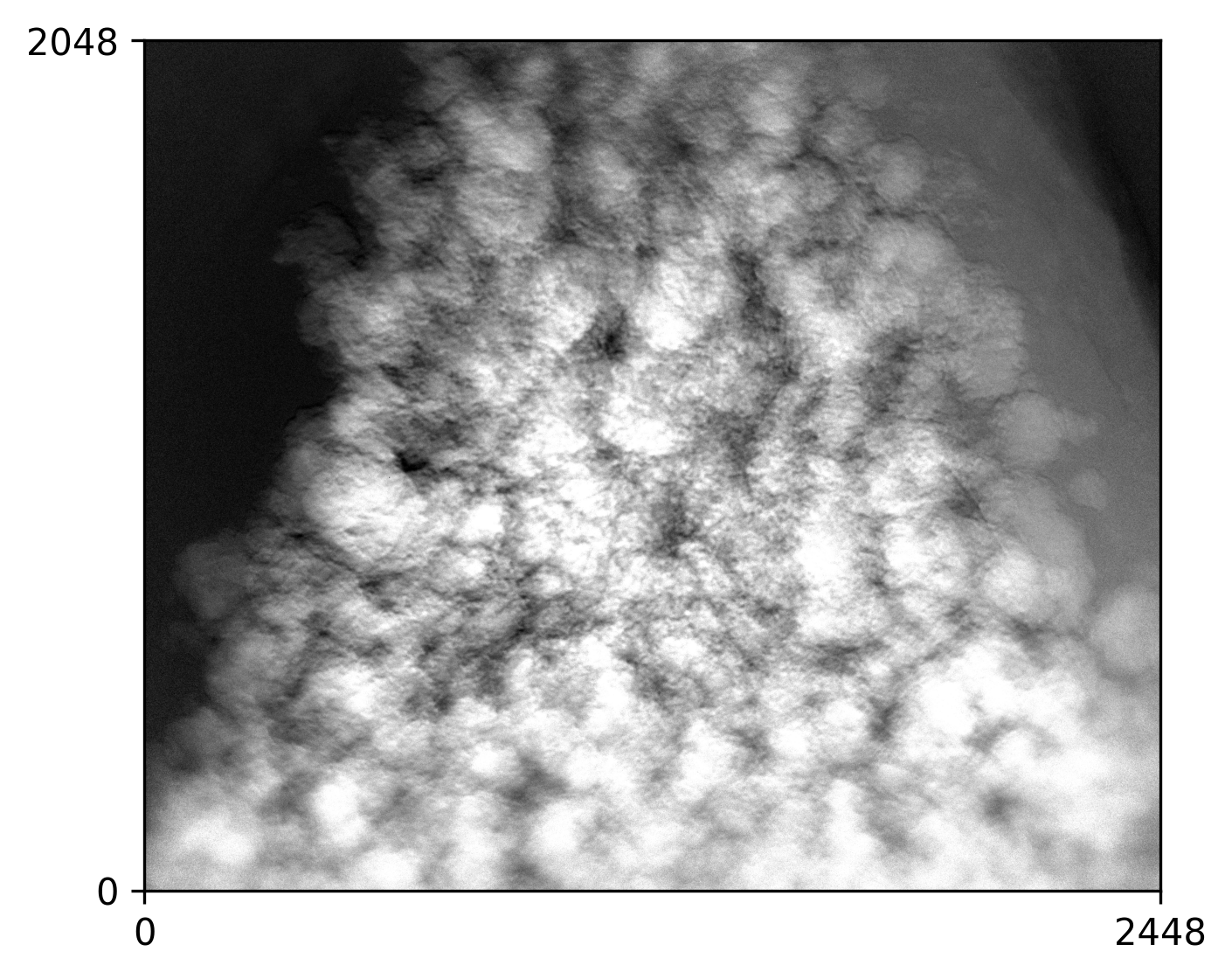}
    }
    \subfigure[]{
     \centering
     \includegraphics[width=0.21\textwidth]{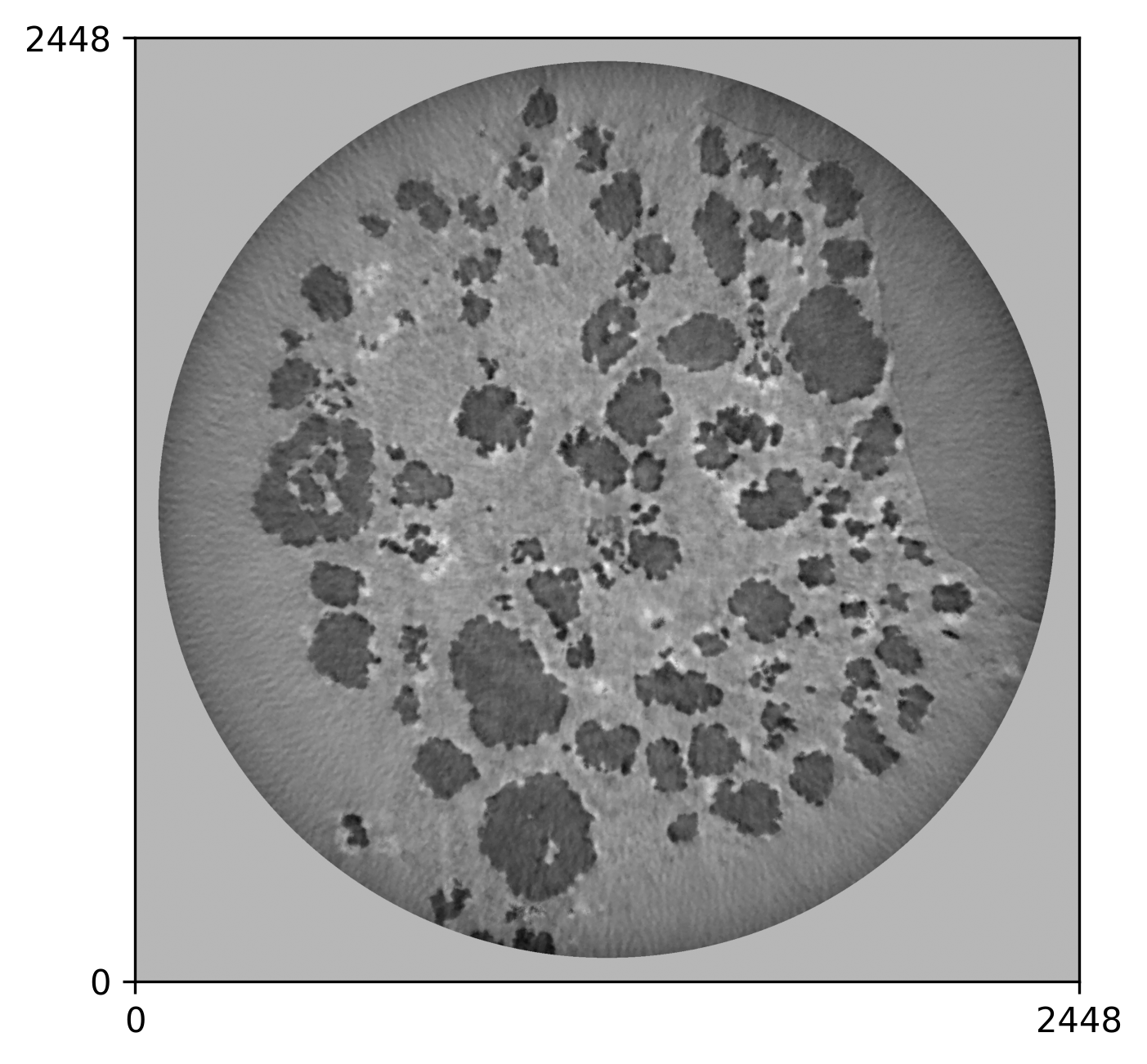}
    }
 \caption{A Nano-CT projection data from the Tomobank was used for benchmarking. (a) 
 A projection image from the tomogram.  (b) A slice of reconstructed volume using MBIR.}
 \label{fig:proj-recon}
\end{figure}

We used a sparsely sampled nano-CT experimental dataset, published on the Tomobank data 
repository \cite{tomobank2018}, to evaluate the performance of the proposed
methods. The dataset consists of $212$ projections with detector 
size $2048 \times 2448$ pixels; see Figure \ref{fig:proj-recon}.
The benchmarks are designed to assess four key aspects of the method: 
(i) the computational
efficiency of the forward and adjoint operators, 
(ii) the impact of the initialization strategy, 
(iii) the effectiveness of the
multi-resolution reconstruction scheme, and 
(iv) the scalability of
the distributed HPC implementation.

\subsection{Toeplitz Acceleration of Forward and Adjoint Operators}

The performance of the forward and adjoint operators is critical
for model-based iterative reconstruction since these operators
dominate the computational cost of each iteration. We compare two
implementations of the operator, i.e., the direct method that uses
$\|R\,f - g\|_2$ and $R^*(R\,f - g)$ to calculate loss and gradient respectively, 
against the reformulated operators that leverage multi-level Toeplitz 
structure \eqref{eq:toeploss} and  \eqref{eq:toepgrad}, across increasing image widths. 

\begin{figure}[h!]
    \centering
    \includegraphics[width=0.4\textwidth]{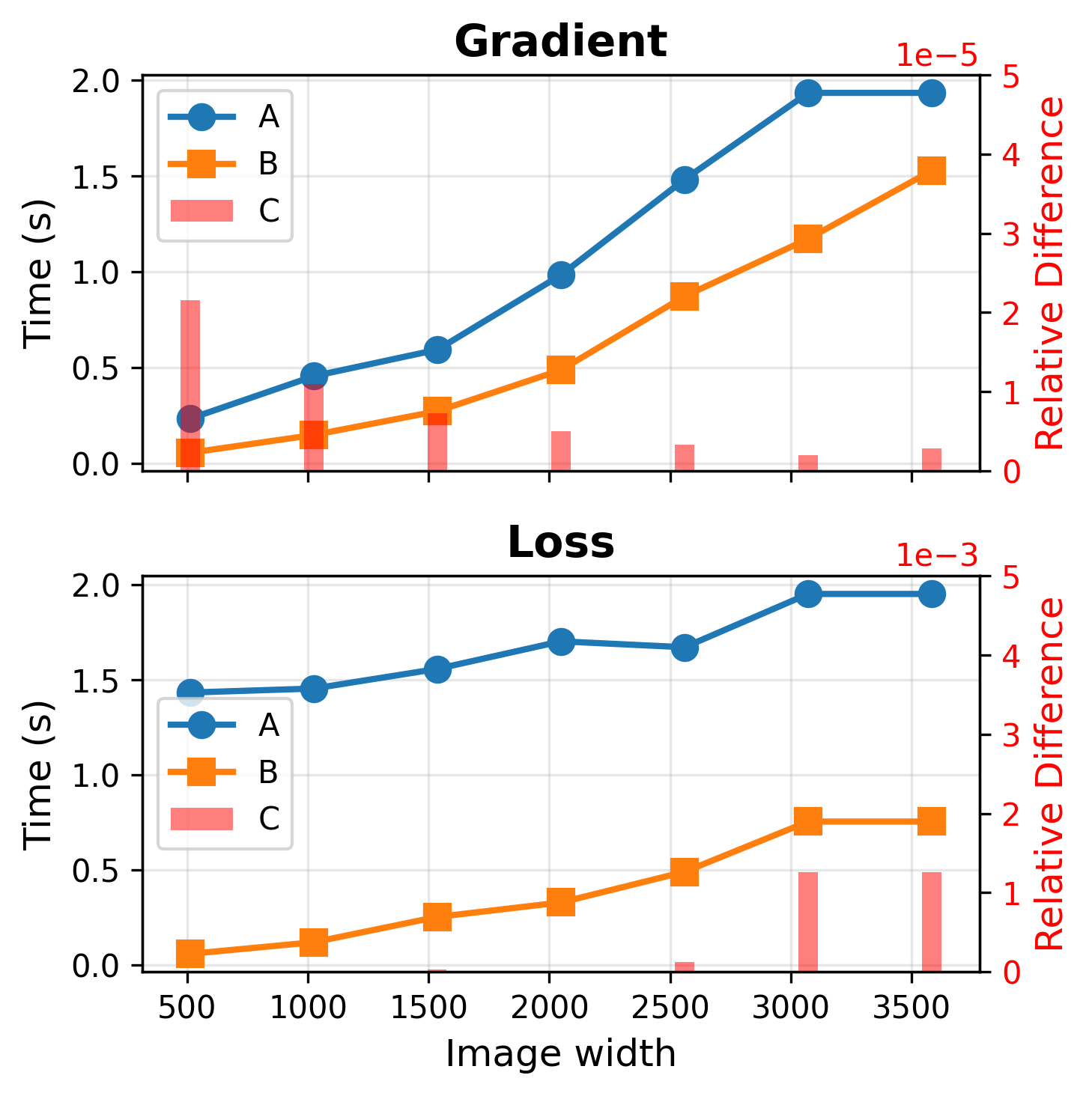}
    \caption{Runtime comparison between the direct method (A) and the Toeplitz-based method (B) for gradient and loss, as well as the relative difference (C) between the two methods, as a function of image width. The optimized Toeplitz implementation consistently 
    loss, as well as the relative difference (C) between the two methods, as a function of image width. The optimized Toeplitz implementation consistently 
    reduces computation time while maintaining high numerical accuracy. The left y-axis represents runtime, 
    and the right y-axis represents the relative difference between the direct and Toeplitz-based methods.}
    \label{fig:toepgrad}
\end{figure}

Figures \ref{fig:toepgrad} report the runtime
for evaluating the gradient and loss. The optimized
implementation consistently reduces computation time while maintaining high numerical accuracy. The relative difference between the two implementations remains small across all tested image sizes, indicating that the
accelerated operator preserves the accuracy of the original method.
The apparent increase in error with image size arises from floating-point accumulation 
effects as more values are summed; however, the maximum deviation remains within 
single-precision (32-bit) machine accuracy of $10^{-6}$ and is therefore negligible in practice.
Overall, these results demonstrate that the optimized operator provides a
significant performance improvement without compromising reconstruction
fidelity.

\subsection{Effect of Initialization}

\begin{figure}[h!]
\centering
\includegraphics[height=4.0cm, width=6.5cm, keepaspectratio=False]{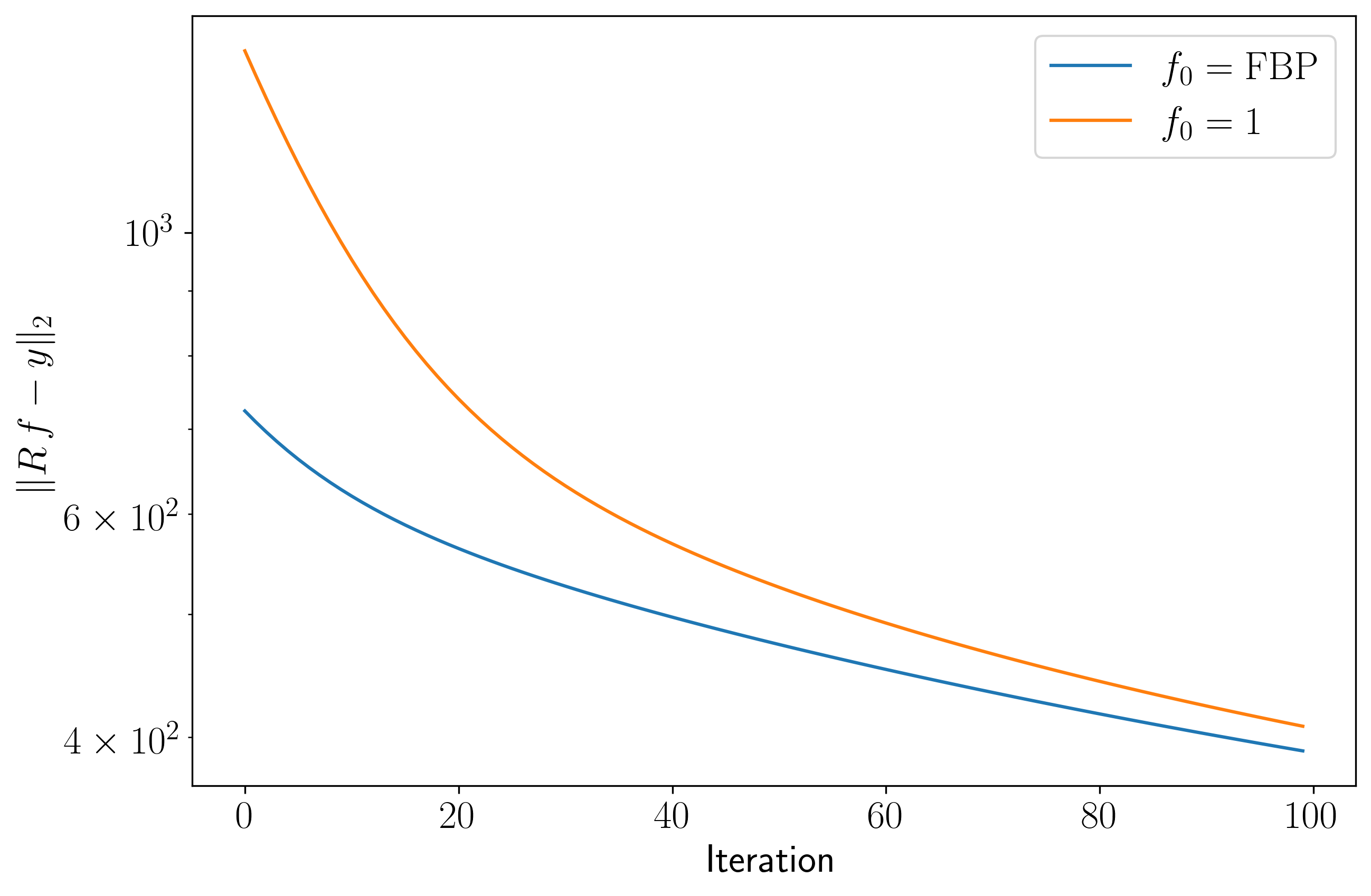}
\caption{The initial residual is approximately \textbf{a factor of two lower} compared to constant initialization, effectively reducing the optimization effort by \textbf{20–25 iterations}. }
\label{fig:init}
\end{figure}

We next evaluate the impact of the proposed filtered backprojection
(FBP) initialization compared with a constant initialization.
Figure~\ref{fig:init} shows the data fidelity term
$\|R\,f - g\|_2$ as a function of iteration for both approaches.

The FBP initialization begins with a substantially lower residual and maintains a consistent advantage throughout the optimization.
This occurs because the FBP estimate captures the dominant low-frequency structure of the object, placing the iterative solver loser to the optimal solution. Consequently, the reconstruction algorithm requires fewer iterations to reach the same residual level compared with a constant initialization. FBP-based initialization significantly reduces the initial residual. 

\subsection{Hierarchical Multi-Resolution Reconstruction}

We now evaluate the impact of the hierarchical multi-resolution
reconstruction strategy. While the filtered backprojection (FBP)
initialization described in the previous subsection naturally serves
as an effective starting point for the coarse grid, it is intentionally
disabled in this experiment to study the effect of the hierarchical reconstruction strategy independently.

\begin{figure}[h!]
\centering
\includegraphics[height=4.0cm, width=6.5cm, keepaspectratio=False]{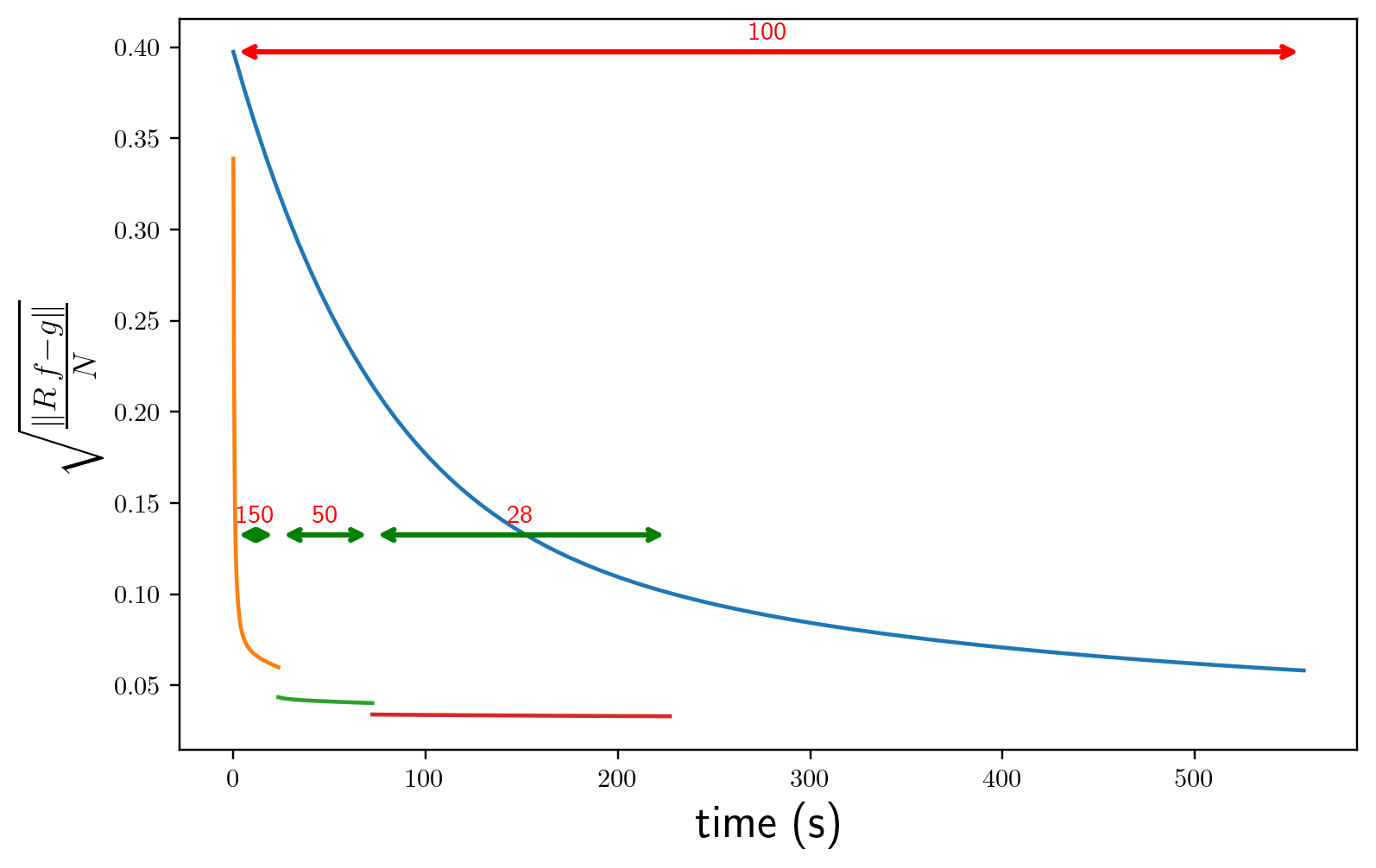}
\caption{Convergence behavior of the multi-resolution reconstruction strategy. The coarse-to-fine approach 
reduces computational cost at the highest resolution by providing improved initial estimates at each refinement stage. 
In this example, three hierarchical grid levels are used, starting from a grid of size ($N /4 \times N/4$), doubling the resolution at each stage. 
Numbers next to arrows indicate iterations at each resolution.}
\label{fig:multires}
\end{figure}

Figure~\ref{fig:multires} compares the
convergence behavior of the proposed coarse-to-fine approach
against a single-resolution optimization. The reconstruction begins on a coarse grid of 
$(611\times611)$, where large-scale structures are recovered efficiently at low computational cost. 
The solution is then interpolated onto progressively finer grids, with the 
resolution increased by a factor of 2 at each stage, providing improved initialization for subsequent optimization.
As shown in Figure~\ref{fig:multires}, this strategy substantially reduces the number of iterations 
required at the highest resolution. While the exact performance gains may vary across datasets, 
in this case the total runtime is reduced by approximately a factor of $5$. 
The coarse-to-fine scheme therefore improves convergence while reducing the overall computational effort.

\subsection{Distributed HPC Performance}

Finally, we evaluate the scalability of the distributed implementation on the Perlmutter supercomputer at the National
Energy Research Scientific Computing Center (NERSC). The
reconstruction volume size in this experiment is
$2048 \times 2447 \times 2447$ voxels.

\begin{figure}[h!]
\centering
\includegraphics[height=4.0cm, width=6.5cm, keepaspectratio=False]{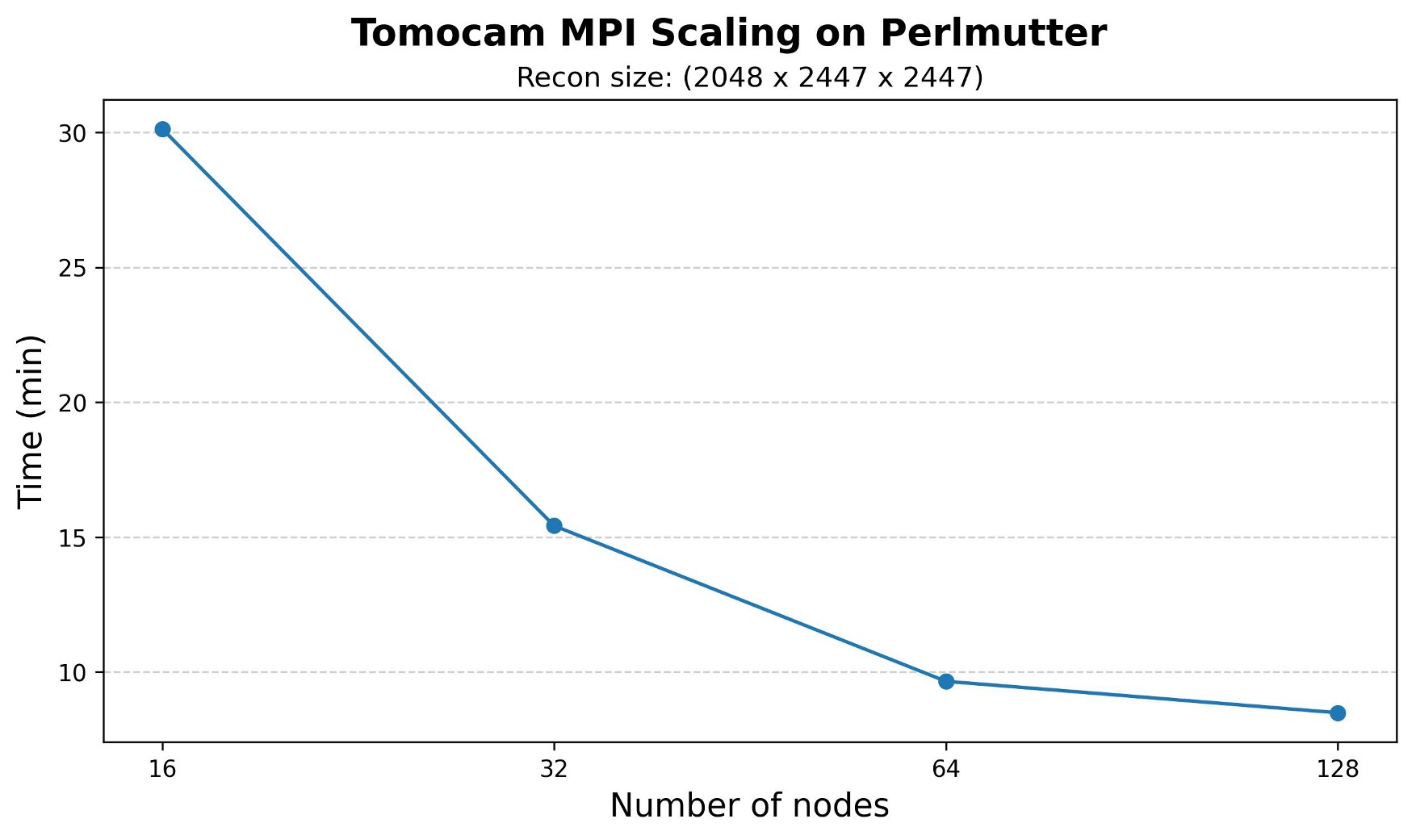}
\caption{Performance of the distributed MPI
implementation on the National Energy Research Scientific Computing Center's Perlmutter supercomputer for a reconstruction volume of size $2048 \times 2447 \times 2447$.}
\label{fig:hpc}
\end{figure}

Figure~\ref{fig:hpc} shows the reconstruction time as a function
of the number of compute nodes. The results demonstrate scaling as the number of nodes increases from 16 to 128. The total reconstruction time decreases from approximately
30 minutes on 16 nodes to less than 9 minutes using 128 nodes. These results confirm that the distributed MPI implementation
effectively utilizes multiple GPUs while keeping communication overhead manageable. The proposed framework therefore enables
efficient reconstruction of large volumetric datasets on modern
HPC systems.

\section{Conclusion}
We presented an efficient framework for large-scale model-based iterative reconstruction (MBIR). By leveraging the multi-level Toeplitz structure of the Radon transform and its adjoint, we reformulated the data fidelity term, enabling significant computational performance improvements.  To further accelerate convergence, we introduced a smart initialization strategy based on filtered backprojection and a hierarchical multi-resolution strategy. This approach captures the low-frequency structure of the object early in the optimization process, leading to faster and more stable convergence. 
We also developed a distributed multi-GPU implementation that partitions the volume along the rotation axis and efficiently manages inter-node inter-GPU communication through halo exchanges. 

Numerical experimental results demonstrate that the proposed strategies achieve significantly faster reconstruction while maintaining quality for large volumetric datasets. These results highlight the practical viability of combining operator-efficient formulations, hierarchical optimization, and distributed computing for next-generation tomographic reconstruction. The presented framework opens the door to near-real-time MBIR for applications such as in situ, in operando, and time-evolving experiments.

\section{Acknowledgments}
This work was supported by the Center for Advanced Mathematics for Energy Research Applications (CAMERA), funded by the Advanced Scientific Computing Research (ASCR) and Basic Energy Sciences (BES) programs of the Office of Science of the U.S. Department of Energy (DOE), as well as by the DOE ASCR Early Career Research Program, under Contract No. DE-AC02-05CH11231.

\bibliographystyle{abbrv}
\bibliography{tomocam_hier}

@book{CCroke_Chap6,
author = {Epstein, Charles L.},
title = {Introduction to the Mathematics of Medical Imaging, Second Edition},
year = {2007},
isbn = {089871642X},
edition = {second}
}

@article{Fessler2003,
author = {Fessler, Jeffrey and Sutton, Brad},
year = {2003},
month = {03}, 
pages = {560 - 574},
title = {Nonuniform Fast Fourier Transforms Using Min-Max Interpolation},
volume = {51},
journal = {Signal Processing, IEEE Transactions on},
doi = {10.1109/TSP.2002.807005}
}

@article{Greengard2004,
author = {Greengard, Leslie and Lee, June-Yub},
title = {Accelerating the Nonuniform Fast Fourier Transform},
journal = {SIAM Review},
volume = {46},
number = {3},
pages = {443-454},
year = {2004},
doi = {10.1137/S003614450343200X}
}

@INPROCEEDINGS{Aditya2014,
  author={Mohan, K. Aditya and Venkatakrishnan, S. V. and Drummy, Lawrence F. and Simmons, Jeff and Parkinson, Dilworth Y. and Bouman, Charles A.},
  booktitle={2014 IEEE International Conference on Acoustics, Speech and Signal Processing (ICASSP)}, 
  title={Model-based iterative reconstruction for synchrotron X-ray tomography}, 
  year={2014},
  pages={6909-6913},
  doi={10.1109/ICASSP.2014.6854939}
 }

@inproceedings{Venkat2013,
	Author = {Singanallur V. Venkatakrishnan and Lawrence F. Drummy and Marc De Graef and Jeff P. Simmons and Charles A. Bouman},
	Booktitle = {Computational Imaging XI},
	Doi = {10.1117/12.2013228},
	Editor = {Charles A. Bouman and Ilya Pollak and Patrick J. Wolfe},
	Organization = {International Society for Optics and Photonics},
	Pages = {75 -- 86},
	Publisher = {SPIE},
	Title = {{Model based iterative reconstruction for Bright Field electron tomography}},
	Url = {https://doi.org/10.1117/12.2013228},
	Volume = {8657},
	Year = {2013}
}

@book{Nesterov2014,
author = {Nesterov, Yurii},
title = {Introductory Lectures on Convex Optimization: A Basic Course},
year = {2014},
isbn = {1461346916},
publisher = {Springer Publishing Company, Incorporated},
edition = {1}
}

@article{tomobank2018,
	doi = {10.1088/1361-6501/aa9c19},
	url = {https://doi.org/10.1088/1361-6501/aa9c19},
	year = 2018,
	month = {feb},
	publisher = {{IOP} Publishing},
	volume = {29},
	number = {3},
	pages = {034004},
	author = {Francesco De Carlo and Do{\u{g}}a Gürsoy and Daniel J Ching and K Joost Batenburg and Wolfgang Ludwig and Lucia Mancini and Federica Marone and Rajmund Mokso and Daniël M Pelt and Jan Sijbers and Mark Rivers},
	title = {{TomoBank}: a tomographic data repository for computational x-ray science},
	journal = {Measurement Science and Technology}
}

@inproceedings{GiselssonB14b,
  author    = {Pontus Giselsson and
               Stephen P. Boyd},
  title     = {Monotonicity and restart in fast gradient methods},
  booktitle = {53rd {IEEE} Conference on Decision and Control, {CDC} 2014, Los Angeles,
               CA, USA, December 15-17, 2014},
  pages     = {5058--5063},
  publisher = {{IEEE}},
  year      = {2014},
  url       = {https://doi.org/10.1109/CDC.2014.7040179},
  doi       = {10.1109/CDC.2014.7040179},
  bibsource = {dblp computer science bibliography, https://dblp.org}
}

@article{RamLak1971,
author={Ramachandran G.N.  and Lakshminarayanan A.V.},
title={{Three-dimensional reconstruction from radiographs and electron micrographs: application of convolutions instead of Fourier transforms}},
journal={{Proc Natl Acad Sci U S A}},
year= {1971},
doi={10.1073/pnas.68.9.2236}
}

@inproceedings{ALS832,
author = {A. A. MacDowell and D. Y. Parkinson and A. Haboub and E. Schaible and J. R. Nasiatka and C. A. Yee and J. R. Jameson and J. B. Ajo-Franklin and C. R. Brodersen and A. J. McElrone},
title = {{X-ray micro-tomography at the Advanced Light Source}},
volume = {8506},
booktitle = {Developments in X-Ray Tomography VIII},
editor = {Stuart R. Stock},
organization = {International Society for Optics and Photonics},
publisher = {SPIE},
pages = {850618},
year = {2012},
doi = {10.1117/12.930243},
URL = {https://doi.org/10.1117/12.930243}
}

@article{APS2BM,
author = "Nikitin, Viktor and Tekawade, Aniket and Duchkov, Anton and Shevchenko, Pavel and De Carlo, Francesco",
title = "{Real-time streaming tomographic reconstruction with on-demand data capturing and 3D zooming to regions of interest}",
journal = "Journal of Synchrotron Radiation",
year = "2022",
volume = "29",
number = "3",
pages = "816--828",
month = "May",
doi = {10.1107/S1600577522003095},
url = {https://doi.org/10.1107/S1600577522003095}
}

@article{NSLS2TXM,
author = {Ge,Mingyuan  and Coburn,David Scott  and Nazaretski,Evgeny  and Xu,Weihe  and Gofron,Kazimierz  and Xu,Huijuan  and Yin,Zhijian  and Lee,Wah-Keat },
title = {One-minute nano-tomography using hard X-ray full-field transmission microscope},
journal = {Applied Physics Letters},
volume = {113},
number = {8},
pages = {083109},
year = {2018},
doi = {{10.1063/1.5048378}},
URL = {{https://doi.org/10.1063/1.5048378}}
}

@article{Larson2018,
title = {Insights from in-situ X-ray computed tomography during axial impregnation of unidirectional fiber beds},
journal = {Composites Part A: Applied Science and Manufacturing},
volume = {107},
pages = {124-134},
year = {2018},
issn = {1359-835X},
doi = {https://doi.org/10.1016/j.compositesa.2017.12.024},
url = {https://www.sciencedirect.com/science/article/pii/S1359835X17304645},
author = {Natalie M. Larson and Frank W. Zok}
}

@article{Kulkarni2020,
doi = {10.1088/2515-7655/abb783},
url = {https://dx.doi.org/10.1088/2515-7655/abb783},
year = {2020},
month = {oct},
publisher = {IOP Publishing},
volume = {2},
number = {4},
pages = {044005},
author = {Devashish Kulkarni and Stanley J Normile and Liam G Connolly and Iryna V Zenyuk},
title = {Development of low temperature fuel cell holders for Operando x-ray micro and nano computed tomography to visualize water distribution},
journal = {Journal of Physics: Energy}
}

@article{tomocam2024,
	title = {{tomoCAM}: fast model-based iterative reconstruction via {GPU} acceleration and non-uniform fast {Fourier} transforms},
	volume = {31},
	url = {https://doi.org/10.1107/S1600577523008962},
	doi = {10.1107/S1600577523008962},
	number = {1},
	journal = {Journal of Synchrotron Radiation},
	author = {Kumar, Dinesh and Parkinson, Dilworth Y. and Donatelli, Jeffrey J.},
	month = jan,
	year = {2024},
	pages = {85--94},
}

@article{boutchko_imaging_2012,
	address = {Netherlands},
	title = {Imaging and modeling of flow in porous media using clinical nuclear emission tomography systems and computational fluid dynamics.},
	volume = {76},
	issn = {0926-9851 1879-1859},
	doi = {10.1016/j.jappgeo.2011.10.003},
	language = {eng},
	journal = {J Appl Geophy},
	author = {Boutchko, Rostyslav and Rayz, Vitaliy L. and Vandehey, Nicholas T. and O'Neil, James P. and Budinger, Thomas F. and Nico, Peter S. and Druhan, Jennifer L. and Saloner, David A. and Gullberg, Grant T. and Moses, William W.},
	month = jan,
	year = {2012},
	pages = {74--81},
}

@article{Marcus:22,
author = {Matthew A. Marcus},
journal = {Opt. Express},
number = {22},
pages = {39445--39465},
publisher = {Optica Publishing Group},
title = {Information content of and the ability to reconstruct dichroic X-ray tomography and laminography},
volume = {30},
month = {Oct},
year = {2022},
url = {https://opg.optica.org/oe/abstract.cfm?URI=oe-30-22-39445},
doi = {10.1364/OE.462410}
}

@article{FESSLER2007191,
title = {On NUFFT-based gridding for non-Cartesian MRI},
journal = {Journal of Magnetic Resonance},
volume = {188},
number = {2},
pages = {191-195},
year = {2007},
issn = {1090-7807},
doi = {https://doi.org/10.1016/j.jmr.2007.06.012},
url = {https://www.sciencedirect.com/science/article/pii/S1090780707002054},
author = {Jeffrey A. Fessler},
}

@mastersthesis{Ou:EECS-2017-90,
    Author= {Ou, Teresa},
    Title= {gNUFFTW: Auto-Tuning for High-Performance GPU-Accelerated Non-Uniform Fast Fourier Transforms},
    School= {EECS Department, University of California, Berkeley},
    Year= {2017},
    Month= {May},
    Url= {http://www2.eecs.berkeley.edu/Pubs/TechRpts/2017/EECS-2017-90.html},
    Number= {UCB/EECS-2017-90},
}

@book{kak2001principles,
  address = {Philadelphia},
  author = {Kak, Avinash C. and Slaney, Malcolm},
  description = {Principles of Computerized Tomographic Imaging (Classics in Applied Mathematics): Aninash C. Kak, Malcolm Slaney: 9780898714944: Amazon.com: Books},
  isbn = {089871494X 9780898714944},
  publisher = {Society for Industrial and Applied Mathematics},
  refid = {46320986},
  title = {Principles of computerized tomographic imaging},
  url = {http://www.amazon.com/Principles-Computerized-Tomographic-Classics-Mathematics/dp/089871494X},
  year = 2001
}

@article{wang2013fourier,
  title={A Fourier-based approach for iterative 3D reconstruction from cryo-EM images},
  author={Wang, Lanhui and Shkolnisky, Yoel and Singer, Amit},
  journal={arXiv preprint arXiv:1307.5824},
  year={2013}
}

\end{document}